\documentclass[conference]{IEEEtran}
\IEEEoverridecommandlockouts
\usepackage{cite}
\usepackage{amsmath,amssymb,amsfonts}
\usepackage{algorithmic}
\usepackage{algorithm2e}
\usepackage{graphicx}
\usepackage{textcomp}
\usepackage{xcolor}
\usepackage{cite}
\usepackage{amsmath,amssymb,amsfonts}
\usepackage{algorithmic}
\usepackage{graphicx}
\usepackage{textcomp}
\usepackage{xcolor}
\usepackage{listings} 
\usepackage{algorithm2e}
\usepackage{booktabs}
\usepackage{multirow}
\usepackage{caption}
\usepackage{subcaption}

\definecolor{darkgreen}{RGB}{0, 153, 51}
\begin{document}

\sloppy

\title{SOLANET: Distributed Neighbor Graph Construction on GPU-Accelerated Systems}

\author{\IEEEauthorblockN{
    Keita Iwabuchi\IEEEauthorrefmark{1},
    Trevor Steil\IEEEauthorrefmark{1},
    Benjamin W. Priest\IEEEauthorrefmark{1},
    Grace Li\IEEEauthorrefmark{1},
    Geoffrey Sanders\IEEEauthorrefmark{1},
    Roger Pearce\IEEEauthorrefmark{1}}
  \IEEEauthorblockA{
    \IEEEauthorrefmark{1}
    Center for Applied Scientific Computing,
    Lawrence Livermore National Laboratory\\
    Livermore, United States of America\\
    \{kiwabuchi, steil1, priest2, li85, sanders29, rpearce\}@llnl.gov}
  \thanks{This material is based upon work supported by the U.S. Department of Energy, Office of Science, Office of Advanced Scientific Computing Research under the project RAMSHORN: Randomized Algorithms for Massive, Scientific, Higher-Order Relational Networks, Award Numbers DE-SCL0000093, SCW1937. This work was performed under the auspices of the U.S. Department of Energy by Lawrence Livermore National Laboratory under Contract DE-AC52-07NA27344 (LLNL-CONF-2017995).}
}


\maketitle

\begin{abstract}
  Neighbor graphs capture relationships among data points and are widely used in data analytics and AI workloads. Many studies have explored approximate construction methods for single-node systems, including GPUs.
  However, extending this to distributed systems for larger data and further acceleration remains challenging due to irregular computation patterns.

  We present SOLANET, a GPU-accelerated distributed neighbor graph construction toolkit. SOLANET first constructs local graphs on each GPU after data partitioning and then refines them via approximate nearest neighbor (ANN) searches over remote graphs pulled from other GPUs using MPI one-sided operations. SOLANET also provides a lock-free single-GPU neighbor graph construction algorithm for AMD GPUs.

  Our single-GPU implementation outperforms a state-of-the-art GPU-based approximate neighbor graph construction implementation across multiple datasets on a single MI300A APU\@. Furthermore, SOLANET demonstrates 11× speedup from 32 to 512 APUs for 1 billion data points and 6.9x speedup from 64 to 512 APUs for 2 billion points.
\end{abstract}


\section{Introduction}
\label{sec:intro}

Given feature vectors and a similarity measure, a neighbor graph captures similarity relationships among points as a graph.
However, constructing exact neighbor graphs, where each point connects to its true nearest neighbors, is expensive in practice.
Thus, approximate neighbor graphs (hereafter simply \emph{neighbor graphs}) are widely used in data analysis.

By traversing the graph, we can find points similar to a query point efficiently.
This task is called approximate nearest neighbor (ANN) search and is used in various applications, such as recommendation engines~\cite{ADENIYI201690}, anomaly detection~\cite{LIAO2002}, and large language models (LLM)~\cite{RAG}.
It also serves as a core component of vector databases~\cite{Pan2024}.
Neighbor graphs are also useful for tasks such as clustering~\cite{1704843, BRITO199733, QIN20181}, dimensionality reduction~\cite{UMAP}, and bioinformatics~\cite{hie2019efficient, wolf2018scanpy, dries2021giotto, van2014accelerating}.

Constructing high-quality neighbor graphs, however, is computationally expensive and often memory intensive.
Moreover, real-world datasets are rapidly increasing in both the number of points and feature dimensionality.
Some datasets already contain billions of points~\cite{BigANNBenchmarks} or have hundreds to tens of thousands of dimensions~\cite{MovieLens}.
Such datasets are often too large for a single processor or machine, or they require more compute than one processor or node can provide in practical time.

Recent large-scale systems increasingly rely on graphics processing units (GPUs) for performance.
Many studies have explored graph-based ANN methods on a single GPU or one multi-GPU machine, such as~\cite{SONG, GPU-NNDescent-Wang, GGNN, CAGRA, BANG, liu2026} (see Section~\ref{sec:related-work}).
However, relatively few studies have addressed distributed systems.
This may be because graph processing approaches typically exhibit irregular computation and communication patterns, making high performance on distributed GPU systems difficult.

To address these challenges, we propose a distributed multi-GPU method for neighbor graph construction.
The design considers both runtime and GPU-system communication characteristics.
After data partitioning, our method first constructs local neighbor graphs on each GPU\@.
It then pulls remote graphs using Message Passing Interface (MPI) one-sided operations and performs local ANN search to find better neighbors residing in other partitions.
To improve its scalability, it uses a binary-tree-structured graph merge scheme.
The distributed model is decoupled from the local graph construction and ANN search algorithms, allowing reuse of existing methods.
We provide a runtime analysis to characterize the model's performance and scalability.

We implement the proposed method for the AMD MI300A APU and evaluate it on the Tuolumne supercomputer at Lawrence Livermore National Laboratory (LLNL), a sibling system of El Capitan with the same architecture.
We call our toolkit SOLANET (\underline{S}calability \underline{O}ptimized \underline{L}arge-scale \underline{A}pproximate \underline{N}eighbor \underline{E}xploration \underline{T}oolkit).
SOLANET also provides a lock-free single-GPU neighbor graph construction algorithm for AMD GPUs.

We first show that our lock-free single-GPU neighbor graph construction implementation outperforms a state-of-the-art GPU-based library on a single MI300A APU while maintaining high graph quality.
We then demonstrate the performance of SOLANET on datasets with up to 2 billion points.
SOLANET achieves 11.0x and 11.7x speedups on two 1-billion-point datasets when scaling from 32 to 512 GPUs, and a 6.9x speedup on a 2-billion-point dataset when scaling from 64 to 512 GPUs.

The main contributions of this paper are as follows:
\begin{itemize}
  \item We propose a distributed neighbor graph construction model, together with a runtime analysis, that is carefully designed to exploit the computation and communication characteristics of distributed GPU systems.
  \item We develop SOLANET, a distributed GPU-based neighbor graph construction toolkit that combines local graph construction with ANN search-based cross-partition refinement to support large-scale datasets efficiently.
  \item SOLANET also provides a lock-free single-GPU neighbor graph construction implementation for AMD GPUs that outperforms a state-of-the-art GPU-based library.
  \item We demonstrate strong scalability on the Tuolumne supercomputer with AMD MI300A APUs for billion-scale datasets, achieving substantial runtime reductions while preserving high graph quality.
\end{itemize}

\section{Preliminaries}
\label{sec:preliminaries}

\subsection{Notation and Problem Setting}
\label{sec:notation}

Let $D$ be a dataset containing $N = |D|$ points (feature vectors).
We assume that every feature vector has the same dimensionality.\footnote{Although our current implementation does not support variable-length feature vectors, doing so would not conflict with the fundamental design of the algorithm. However, it may require some engineering effort to support variable-length vectors efficiently for GPU, and we leave this for future work.}

\begin{figure}
  \centering
  \includegraphics[width=1.0\columnwidth]{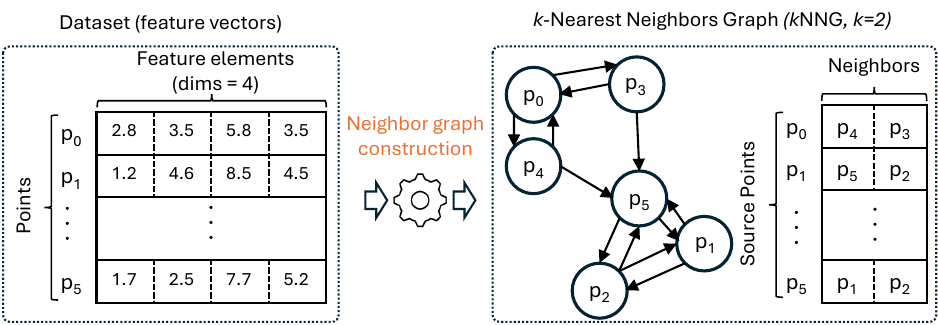}
  \caption{Overview of nearest neighbor graph construction.}
  \label{fig:nng-overview}
\end{figure}

The goal of this work is to construct a high-quality neighbor graph for a given dataset $D$ at massive scales.
We focus on constructing an \emph{approximate $k$-nearest neighbor graph} ($k$NNG), in which each source point has exactly $k$ approximate nearest neighbors ($1 \le k$, and we assume $k \ll N$).
This simple structure is easy to use in downstream applications.
We use $G$ to denote a $k$NNG, and define the quality of $G$ by its recall, namely, the percentage of neighbors in $G$ that also appear in the exact $k$NNG computed by brute force.
Fig.~\ref{fig:nng-overview} illustrates an overview of the neighbor graph construction task.

In this work, the $k$NNG is represented as a 2D matrix of size $N \times k$, allocated as a contiguous row-major memory block.
Each row contains the $k$ neighbors of the corresponding source point.
A neighbor entry stores a neighbor point ID and its distance from the source point.
We store neighbor IDs and distances in two separate 2D matrices: $G_{id}$ for neighbor IDs and $G_{dis}$ for distances.

Let $k_s$ denote the number of neighbors to find in a single approximate nearest neighbor (ANN) search for a query point ($1 \le k_s < N$, usually $k_s \ll N$, and can be $k_s \le k$).
$G^s$ denotes a $k$NNG optimized for ANN search.

Our algorithm can use any similarity measure that can compare any pair of data points.
Here, however, we consider only pairwise measures that correspond to proper distance metrics.
We use $\sigma(p_1, p_2)$ to denote the \emph{distance} between $p_1$ and $p_2$, where $p_1, p_2 \in D$.
We assume that the distance function $\sigma$ returns a nonnegative value $d$, where smaller values indicate more similar points.
If the original function returns \emph{similarity} values, we can transform it into a distance-like \emph{dissimilarity} score, as is common in ANN search.

We use Message Passing Interface (MPI) for inter-process communication.
We assume a distributed system with $P$ MPI ranks and a one-to-one correspondence between MPI ranks and GPUs.


\subsection{MI300A Accelerated Processing Units (APUs)}

Although our distributed neighbor graph construction algorithm is designed to be independent of any specific hardware architecture, in this paper we focus on the AMD MI300A Accelerated Processing Unit (APU) architecture.
It is used in the El Capitan supercomputer at Lawrence Livermore National Laboratory (LLNL), which is ranked first on the TOP500 list as of November 2025~\cite{TOP500}.
Its programming environment is based on the AMD ROCm software stack, and HIP is the C++ programming API for GPU programming.

Its most notable feature is that it shares physical memory between CPU and GPU\@.
Once the unified memory mode is enabled, even memory allocated by standard CPU memory allocation functions (e.g., \texttt{malloc}) can be accessed by the GPU without explicit data transfer operations.
However, memory access performance can vary significantly with the allocation API and which processor (CPU or GPU) touched the memory first.
Wahlgren et al.\ studied these memory architecture characteristics of MI300A APUs~\cite{Wahlgren2025}, including the impact of different allocation APIs, and found that the best performance is achieved with \texttt{hipMalloc}.
We use \texttt{hipMalloc} for memory allocation of core data structures in our implementation.



\subsection{Related Work}
\label{sec:related-work}

We focus on prior work most directly related to SOLANET: graph-based ANN on GPUs and other large-scale ANN indexing approaches.

\subsubsection{Graph-Based ANN on Single Node}
Many ANN studies have focused on GPUs within a single machine.
For example, SONG~\cite{SONG}, GNND~\cite{GPU-NNDescent-Wang}, GGNN~\cite{GGNN}, CAGRA~\cite{CAGRA}, BANG~\cite{BANG}, and Faiss~\cite{Faiss} are representative GPU-based libraries for neighbor graph construction.

To leverage these efforts, SOLANET decouples the distributed computation model from the local graph construction and ANN search algorithms.

\subsubsection{Distributed Graph-Based ANN}

Several studies used MapReduce~\cite{MapReduce} frameworks for distributed neighbor graph construction~\cite{nndescent, WARASHINA}.
These methods are based on NN-Descent~\cite{nndescent}, a widely-used, heuristic, iterative algorithm for neighbor graph construction.
However, their performance is limited by the design and implementation of MapReduce-style frameworks.
Kim et al.~\cite{Kim} further improved over~\cite{WARASHINA}; however, their method relies on centroid information, which may not be suitable for some similarity measures, and their paper and implementation support only the $L_2$ distance metric.

DNND~\cite{DNND} and its follow-up work, NEO-DNND~\cite{NEO-DNND}, are CPU-based distributed NN-Descent libraries that use the Message Passing Interface (MPI) for communication.
They send feature vectors to other ranks on demand to strictly follow the original NN-Descent algorithm and perform only one distance calculation per communicated feature vector, which yields low arithmetic intensity and is not suitable for GPU-based systems.
We compare SOLANET with NEO-DNND in our evaluation and discuss the performance differences.

PyRKNN~\cite{PyRKNN} is a distributed neighbor graph construction library that demonstrated its performance on up to 128M points using 64 NVIDIA V100 (16 GB) GPUs.
It partitions the dataset using random projection trees, performs local exact $k$NN searches within each subset, and merges the results until the global approximate graph converges.
Due to its use of random projection trees for partitioning, it is limited to $L_2$ and cosine distance metrics.
In contrast, SOLANET finds cross-partition neighbors by performing ANN searches explicitly on each remote subgraph.
This design makes SOLANET less sensitive to partition quality and agnostic to the choice of similarity measure.
We also demonstrate SOLANET on datasets with up to 2 billion points using 512 MI300A (128 GB) APUs.

Several other works partition the dataset into subsets and construct neighbor graphs independently within each partition, for example in CPU-based distributed settings~\cite{RengaBashyam2020}.
In contrast, SOLANET constructs a globally connected graph by explicitly recovering cross-partition neighbors, which is a more difficult problem.

\subsubsection{Other Related Work}

To handle large datasets that do not fit in a single machine's memory, some ANN frameworks utilize node-local storage.
For example, DiskANN~\cite{DiskANN} is a disk-based ANN framework that can handle billion-scale datasets on a single machine by efficiently utilizing local storage devices.
Because SOLANET does not rely on any specific local graph construction or ANN search algorithm, it could also incorporate this type of approach.

\section{Local $k$NNG Construction and ANN Search}
\label{sec:local-implementation}

Our distributed neighbor graph construction algorithm relies on two single-GPU building blocks: local $k$NNG construction and local approximate nearest neighbor (ANN) search.

It does not require a specialized local $k$NNG construction algorithm: any single-GPU method could be used if it produces high-quality graphs in practice and supports the desired similarity measures.

However, compared to NVIDIA CUDA GPUs, there are few GPU-based $k$NNG construction libraries for AMD GPUs, so we implemented one for a single AMD GPU\@.

For local ANN search, we use CAGRA~\cite{CAGRA}, a state-of-the-art GPU-based ANN search algorithm.
CAGRA is implemented in the cuVS library~\cite{cuVS} and the hipVS library~\cite{hipVS}.
hipVS is a HIP version of cuVS and works on AMD GPUs.

In the following sections, we describe our local $k$NNG construction algorithm and the ANN search background most relevant to the distributed method.

\subsection{NN-Descent}
\label{sec:nndescent}

Our local $k$NNG construction is based on the NN-Descent algorithm~\cite{nndescent}.
It is highly effective at producing high-quality graphs in practice~\cite{Wang} and is one of the most widely used heuristic algorithms for constructing  $k$NNGs.
Its empirical cost is approximately $O(N^{1.14})$~\cite{nndescent}, which is significantly better than the $O(N^2)$ cost of brute-force methods.
It also supports arbitrary distance functions and the final output is a $k$NNG that contains exactly $k$ approximate neighbors per point.

The core idea of NN-Descent is: \textit{if two points have a common neighbor, they are likely to be neighbors of each other}.
It first initializes a $k$NNG ($G$) with randomly selected neighbor points.
It then checks for closer relationships within each point's neighbors in $G$, exploiting the idea that points sharing a common neighbor are likely close to each other.
It repeats this process iteratively until the number of accepted neighbor updates falls below $\delta kN$, where $\delta$ is the convergence threshold (e.g., 0.0001) and $N$ is the number of data points.
To reduce redundant distance calculations, it only considers a subset of neighbors for each point at each iteration, which is controlled by a sampling rate parameter $\rho$.

\subsubsection{Lock-Free GPU-based NN-Descent}
\label{sec:lock-free-nndescent}

We developed a lock-free GPU-based NN-Descent algorithm by adopting several techniques from GNND~\cite{GPU-NNDescent-Wang},
a state-of-the-art GPU-based NN-Descent algorithm used in cuVS~\cite{cuVS} and hipVS~\cite{hipVS}.

The key difference is that our implementation is lock-free, whereas GNND uses locks to update the $k$NNG ($G$).
During the NN-Descent algorithm, multiple threads may try to update the neighbor list of the same point in the $G$ simultaneously.
GNND uses spin locks allocated in the GPU's global memory to prevent race conditions.
However, acquiring and releasing locks in global memory significantly degrades performance.
Therefore, they mitigate the performance degradation by attempting neighbor list updates only for highly prioritized cases.

We aim to further accelerate the neighbor update step by using atomic operations.
Specifically, rather than updating $G$ directly under locks, we maintain a \emph{neighbor candidate list} for each point and atomically append newly found candidates to this list if they are closer than the current farthest neighbor of that point.
Then, we assign a single thread to process the candidate list of each point and update the $k$NNG without worrying about race conditions.
After updating the $k$NNG, we clear the candidate list and move on to the next iteration.

To use GPU memory efficiently, we do not dynamically increase the size of the candidate list; as a result, our method may drop some true neighbors if the list becomes full.
However, the iterative nature of NN-Descent should let the algorithm retry the same neighbor check in later iterations.
We can also tune some NN-Descent parameters to conduct more redundant neighbor checks on purpose.
Our algorithm accepts extra computation to increase parallelism, a common GPU design tradeoff.

We evaluate the performance and accuracy of our lock-free GPU-based NN-Descent implementation in Section~\ref{sec:evaluation}.

\subsection{Utilizing Near-Constant ANN Search Performance}
\label{sec:local-ann-search-performance}

Another important building block of our distributed method is local approximate nearest neighbor (ANN) search.
To find closer neighbors across partitions efficiently, we run ANN search on graphs pulled from other GPUs (see Section~\ref{sec:dist-construction} for the distributed algorithm).

A graph-based ANN search function takes query points, a $k$NNG, the dataset associated with the $k$NNG, and a few algorithm parameters such as $k_s$ (the number of neighbors to find per query point, usually $k_s \ll N$), as inputs.

For the local ANN search, we use the CAGRA~\cite{CAGRA} search implementation in hipVS~\cite{hipVS}.
CAGRA initiates the search from randomly selected points in the $k$NNG and greedily explores the graph by visiting neighbors of previously visited points.
The search continues until no closer neighbors are found or a stopping criterion is satisfied.
While the overall design of the search algorithm is standard, CAGRA is highly optimized for GPU architectures and is among the best-performing GPU-based ANN search implementations.

Our preliminary experiments confirmed an important characteristic of ANN search that strongly influences our distributed design: the runtime of ANN search tends to be largely insensitive to the number of source points in the $k$NNG\@.
For example, ANN searches on $k$NNGs constructed from 1 million points and 10 million points have similar runtime when other conditions are similar (e.g., the same $k$ and search parameters).


We attribute this efficiency to graph-based ANN search itself, which is designed to find promising candidates within a small number of hops from the initial query start points.
Furthermore, neighbor graphs are often optimized for ANN search beforehand to amplify this effect.
Our distributed graph construction algorithm exploits this performance characteristic as described in Section~\ref{sec:scalable-algorithm-overview}.

\subsection{Graph Optimization for ANN Search}
\label{sec:graph-optimization}

To achieve high ANN search performance, it is common to optimize a $k$NNG before using it as the search index.
We implemented the optimization technique proposed by CAGRA~\cite{CAGRA} because it is highly efficient and uses a $k$NNG constructed by NN-Descent.
The idea is based on NSG~\cite{NSG}, and similar optimization is also used in other ANN search libraries such as pyNNDescent~\cite{pynndescent}.

The core idea is to remove redundant edges that can be bypassed via shorter paths.
The goal is to maximize the number of neighbors reachable within a small number of hops from the initial query start points.
CAGRA adapts this optimization to GPU architectures and achieves significant performance improvement during ANN search with minimal graph optimization overhead.
It also does not require distance data in the $k$NNG, which is beneficial for memory and communication efficiency.

We refer to the optimized $k$NNG as a \textit{search graph}, denoted by $G^s$, to distinguish it from the original $k$NNG ($G$).
\section{Distributed GPU-Based Neighbor Graph Construction}
\label{sec:dist-construction}

We now present SOLANET's distributed GPU-based neighbor graph construction algorithm.

Given a dataset $D$, our algorithm randomly partitions the points across $P$ MPI ranks.
The output $k$NNG follows the same partitioning: all $k$ neighbors of a source point are stored on the same rank as the point itself.

After partitioning the data, our algorithm first constructs a local $k$NNG on each rank independently.
Because the computational cost of constructing a $k$NNG grows superlinearly with the number of data points, this step scales well with the number of partitions.
However, these constructed local $k$NNGs do not include neighbors residing on other ranks, so they must be updated to discover better cross-partition neighbors.
We refer to this process as \emph{distributed $k$NNG refinement}.
Our refinement algorithm leverages approximate nearest neighbor (ANN) search to identify better neighbors across partitions.
Each MPI rank performs ANN searches for its local data points on $k$NNGs pulled from other ranks and then updates its local $k$NNG based on the ANN search results.

To fetch remote graphs, we use MPI one-sided communication so that each rank can fetch remote data asynchronously without interrupting the target rank.

In the following subsections, we first describe a straightforward ANN search-based refinement approach and its limitations.
We then present our scalable distributed $k$NNG refinement algorithm and runtime analysis.

\subsection{Notation}

General notation, such as $D$, $G$, $G^s$, $k$, and $k_s$, is introduced in Section~\ref{sec:notation}.
Here, we introduce notation specific to distributed graph construction.

The local dataset and local $k$NNG on rank $i$ are denoted by $D_i$ and $G_{i \rightarrow i}$, respectively, and the corresponding search graph by $G^s_{i \rightarrow i}$.
More generally, $D_{a:b} = D_a \cup \cdots \cup D_b$ denotes the concatenation of dataset partitions from rank $a$ through rank $b$, and $G_{a:b \rightarrow c:d}$ denotes a $k$NNG whose source points are in $D_{a:b}$ and whose neighbors are in $D_{c:d}$.
$R_{a \rightarrow b:c}$ denotes the ANN search results for query points in $D_a$ obtained by searching $G^s_{b:c \rightarrow b:c}$; for each query point, $R_{a \rightarrow b:c}$ contains its approximate nearest neighbors in $D_{b:c}$.
We use $M$ to denote the number of groups remaining after the binary-tree-structured refinement phase.

For communication modeling, we use the Hockney model~\cite{Hockney1994}. In this model, the time to send a message of size $m$ between two processes is $T(m) = \alpha + m\beta$, where $\alpha$ represents latency and start-up cost and $\beta$ is the inverse bandwidth.

\subsection{All-to-All ANN Search-Based Graph Refinement}

To refine local $k$NNGs, a straightforward baseline is an all-to-all ANN search-based refinement, in which each rank performs ANN searches for its local points on the $k$NNGs pulled from all other ranks and updates its local $k$NNG based on the results.
Specifically, each rank executes the following steps independently:

\begin{enumerate}
  \item Construct the local $k$NNG $G_{i \rightarrow i}$ on MPI rank $i$ from the local dataset $D_i$.
  \item Optimize $G_{i \rightarrow i}$ for ANN search to obtain $G^s_{i \rightarrow i}$ using the method from Section~\ref{sec:graph-optimization}, then synchronize all ranks with an \texttt{MPI\_Barrier()} call.
  \item On rank $i$, for a remote rank $j \neq i$, pull $G^s_{j \rightarrow j}$ and the associated dataset $D_j$.
  \item Perform ANN searches for the local points on the pulled $G^s_{j \rightarrow j}$, find $k$ candidate neighbors per point, and update $G_{i \rightarrow i}$ accordingly ($G_{i \rightarrow i}$ becomes $G_{i \rightarrow i:j}$ if $i < j$).
  \item Repeat steps 3--4 for all remote ranks $j$.
\end{enumerate}

This simple algorithm can construct a $k$NNG for datasets beyond a single GPU's memory capacity, but its strong scalability is limited because graph-based ANN search can have near-constant runtime with respect to the number of source points, as discussed in Section~\ref{sec:local-ann-search-performance}.
In this approach, each rank finds neighbor candidates for its $\frac{N}{P}$ local points by searching each of the $P-1$ graphs pulled from the other ranks, where $N$ is the total number of points and $P$ is the number of MPI ranks.
If we assume constant ANN search cost $S$ per query regardless of the number of source points in the search graph, then the refinement compute time is
\[
  S \cdot \frac{N}{P} \cdot (P-1) \approx S \cdot \frac{N}{P} \cdot P = SN,
\]
and the communication time is
\[
  \left(\alpha + \frac{N}{P} \beta \right) \cdot (P-1) \approx P \alpha + N \beta.
\]

Here, $\frac{N}{P}\beta$ denotes, up to constant factors, the bandwidth cost of transferring the full remote payload required for one refinement round, including $G^s$ and $D$.

The latency term of this simple algorithm scales with $P$, and even if we ignore that, the compute and bandwidth terms remain fixed as $P$ increases.
Thus, increasing $P$ does not reduce refinement time, so we need another approach.

\subsection{Scalable Algorithm Overview}
\label{sec:scalable-algorithm-overview}
Our goal is better strong scalability: in many practical settings, the dataset size is fixed, so increasing the number of compute nodes should reduce time-to-solution.

To this end, we propose a two-phase distributed $k$NNG refinement algorithm: (i) binary-tree-structured distributed graph refinement (Section~\ref{sec:binary-tree-refinement}) and (ii) flat distributed $k$NNG refinement (Section~\ref{sec:flat-refinement}).

Phase~(ii) is closely related to the all-to-all ANN search-based refinement described above.
As discussed in Section~\ref{sec:local-ann-search-performance}, ANN search cost is nearly constant in practice with respect to the number of source points in the search graph. Therefore, searching one larger graph is more efficient than searching multiple smaller graphs. Phase~(i) exploits this by merging local $k$NNGs into larger sub-$k$NNGs before phase~(ii), allowing each subsequent ANN search to cover a larger search space per round.

When GPU memory is insufficient for phase~(i), we skip it and directly run phase~(ii) on the local $k$NNGs.

\begin{figure*}
  \centering
  \includegraphics[width=1.0\textwidth]{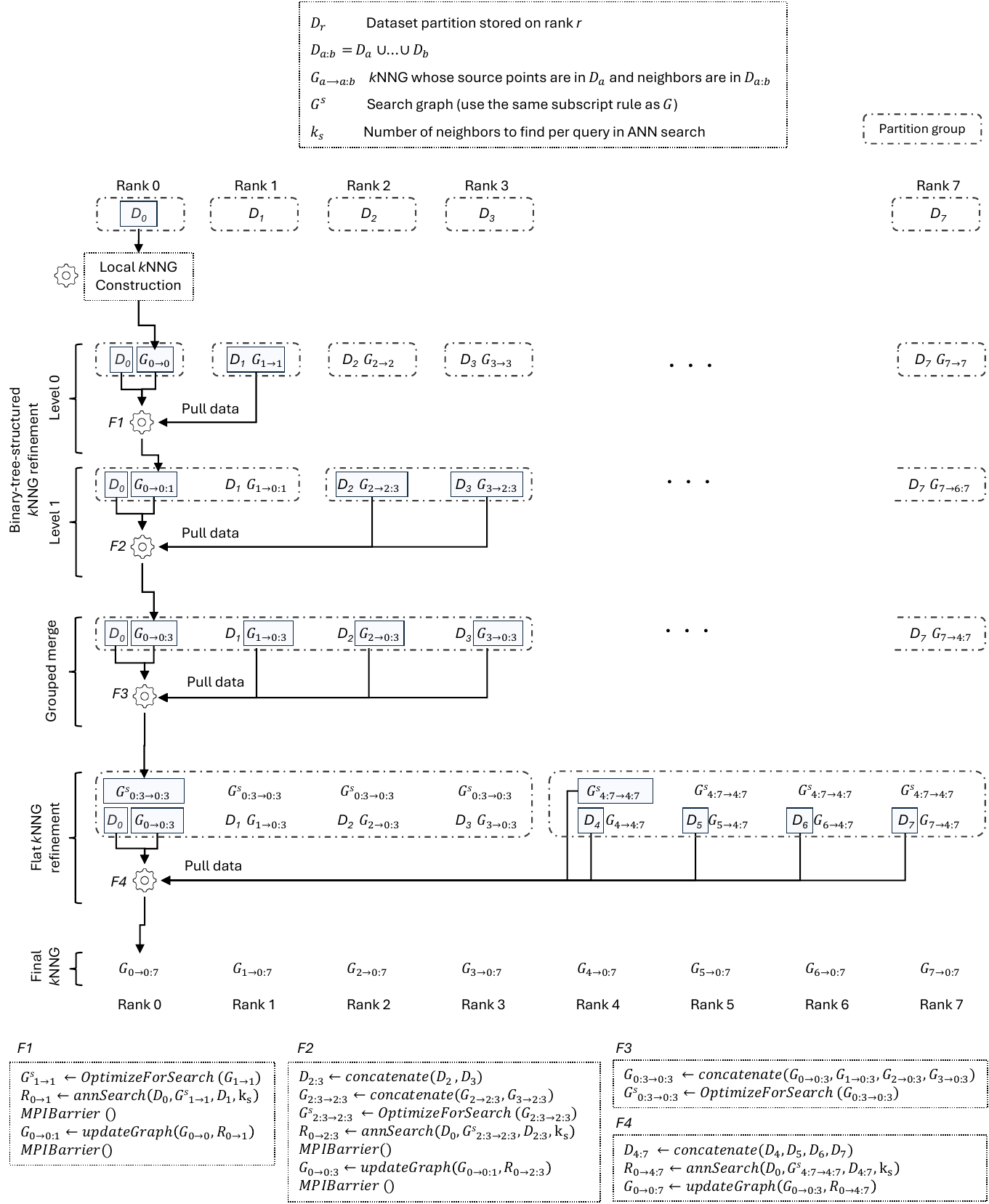}
  \caption{ANN search-based distributed $k$NNG refinement with 8 MPI ranks. Assume that at least two GPUs are required to store the dataset. Only the activity of rank~0 is shown for simplicity. All ranks execute the same procedure.
  }
  \label{fig:dist-knng-refine}
\end{figure*}

Fig.~\ref{fig:dist-knng-refine} presents an overview of the refinement procedure for 8 MPI ranks (one GPU per rank).
For simplicity, we describe only the activity of rank~0, though all ranks execute the same procedure in parallel.

\subsection{Binary-Tree-Structured ANN Search-Based Graph Refinement}
\label{sec:binary-tree-refinement}

We use this binary-tree-structured refinement when there are sufficient GPU resources to store additional data to conduct this technique.

In this phase, local $k$NNGs are merged hierarchically until $M$ groups remain, where $M$ is the target number of groups after the binary-tree-structured phase.
We assume that $M$ is a power of two and that $2 \leq M \leq P$, where $P$ is the total number of MPI ranks.
In Fig.~\ref{fig:dist-knng-refine}, $P=8$ and $M=2$.

At each level, MPI ranks are partitioned into equally sized adjacent \emph{groups} according to the binary tree structure.
The group size at level $l$ is $2^l$.

The $annSearch()$ function in Fig.~\ref{fig:dist-knng-refine} takes four arguments: (1) query points, (2) a search graph, (3) feature vectors associated with the search graph, and (4) $k_s$, the number of neighbors to return per query point ($k_s = k$ by default).
Because the local dataset already has the required 2D matrix layout, it can be used directly as the query matrix.
ANN search results $R$ consist of two 2D arrays: one for neighbor IDs and the other for distances to those neighbors from the query points.

At level~0, rank~0 pairs with rank~1, pulls $D_1$ and $G_{1 \rightarrow 1}$, optimizes the pulled graph to obtain $G^s_{1 \rightarrow 1}$, searches it for neighbor candidates of the points $D_0$, and updates $G_{0 \rightarrow 0}$ to $G_{0 \rightarrow 0:1}$ using $R_{0 \rightarrow 1}$.
At level~1, ranks~0 and~1 are paired with ranks~2 and~3. Rank~0 gathers $D_2$ and $G_{2 \rightarrow 2:3}$ from rank~2 and $D_3$ and $G_{3 \rightarrow 2:3}$ from rank~3, concatenates them, and optimizes the concatenated graph to achieve $G^s_{2:3 \rightarrow 2:3}$.
It then searches for neighbor candidates of points $D_0$ on $G^s_{2:3 \rightarrow 2:3}$ and updates $G_{0 \rightarrow 0:1}$ to $G_{0 \rightarrow 0:3}$ using the search results.
This procedure repeats at higher levels until only $M$ groups remain.

When multiple datasets and $k$NNGs are pulled, they are stored consecutively and treated as one dataset and graph.
Because they are stored as contiguous 2D arrays, no additional reordering is needed, so the $concatenate()$ function in Fig.~\ref{fig:dist-knng-refine} updates only metadata such as the number of rows.

Before each local graph update, we issue a global synchronization operation such as \texttt{MPI\_Barrier} so that no rank is still pulling graph data while another rank modifies it.
After the update, we issue a second barrier so that all ranks complete the current level before proceeding.

\subsection{Merging $k$NNGs within Groups}
\label{sec:grouped-merge}

After the binary-tree-structured refinement phase, we move to the \emph{grouped merge} phase, if more than one rank belongs to each group.
In Fig.~\ref{fig:dist-knng-refine}, rank~0 belongs to the group of ranks 0--3.
In this phase, it gathers the graphs from ranks 1, 2, and 3, concatenates them with its own graph to form $G_{0:3 \rightarrow 0:3}$, and produces a search graph $G^s_{0:3 \rightarrow 0:3}$.
After this phase, each rank has a search graph that covers the dataset of its group.

\subsection{Flat ANN Search-Based Graph Refinement}
\label{sec:flat-refinement}

In the flat $k$NNG refinement phase, each rank performs ANN searches from its local points against the search graphs of all other groups.
Fig.~\ref{fig:dist-knng-refine} illustrates this phase for the case of 2 groups, again showing only rank~0.

At this stage, all ranks within the same group hold the same search graph $G^s$, while the dataset $D$ remains partitioned across ranks to reduce memory use.
A rank in the $i$-th position within its group pulls $G^s$ from the rank in the same position in each of the other groups to maximize communication parallelism.

Continuing the same example, rank~0 pulls the search graph from rank~4 and pulls the datasets from ranks 4--7.
It then uses its local points as queries against that pulled graph and updates $G_{0 \rightarrow 0:3}$.
If there are more than 2 groups ($M > 2$), the same procedure repeats for the search graphs of the other groups.

This phase does not require global synchronization because $G^s$ is read-only, so each rank can proceed independently.
Communication can also be overlapped with computation using double buffering, which partially hides the latency of pulling remote data behind ANN search and graph update.

After the flat refinement phase, all local $k$NNGs $G$ contain approximate nearest neighbors selected from the entire dataset.

\subsection{Summary of Algorithm Advantages}

The proposed distributed $k$NNG refinement algorithm has the following advantages:
\begin{itemize}
  \item It can directly leverage existing literature on graph-based ANN, as the distributed graph refinement model is decoupled from the local ANN $k$NNG construction and search backend.
  \item The algorithm has high arithmetic intensity, that is, a high computation-to-communication ratio, which is favorable for distributed GPU systems.
  \item Its communication pattern is regular and coarse-grained, which helps apply optimizations and utilize network bandwidth efficiently.
  \item Major data transfers use MPI one-sided \emph{get} operations, allowing each rank to fetch remote data without active participation from the target rank.
  \item The transferred data structures ($k$NNGs and datasets) are simple and can be placed directly in GPU-accessible memory after communication, avoiding additional transformation or copying. This can be achieved even on conventional CPU-GPU systems, where memory is not shared, by using GPUDirect RDMA~\footnote{https://developer.nvidia.com/gpudirect}, for example.
\end{itemize}

These advantages contribute to the high performance and scalability demonstrated in Section~\ref{sec:evaluation}.

\subsection{Runtime Analysis}

We now analyze the runtime of the proposed distributed $k$NNG refinement algorithm and explain why it is expected to scale strongly, focusing on the dominant costs.

We make the following assumptions:
(i) the computation uses $P$ MPI ranks, and each rank stores $\frac{N}{P}$ data points;
(ii) all ranks progress in parallel without load imbalance;
(iii) the graph optimization cost is also negligible relative to the ANN search cost; and
(iv) the cost of one ANN search from one query point is a constant $S$, independent of the number of source points in the search graph.
In the communication terms below, $\frac{N}{P} \beta$ similarly denotes the bandwidth cost of the full remote payload needed for one ANN search round, rather than only one specific array.
We ignore other factors, such as the dataset's dimensionality, $k$, and $k_s$, because they are unchanged with respect to the number of MPI ranks.

We first consider the binary-tree-structured refinement phase.
At tree level $i$, each rank performs ANN searches for its $\frac{N}{P}$ local points on a search graph $G^s$ of size $2^{i} \cdot \frac{N}{P}$ that needs to be pulled from $2^{i}$ other ranks, so the runtime per level is
$S \cdot \frac{N}{P} + 2^{i} \left( \alpha + \frac{N}{P} \beta \right)$.
Reducing the number of groups from $P$ to $M$ requires $L = \log_2\!\left(\frac{P}{M}\right)$ levels.
So, the total runtime of this phase is

\begin{align*}
  \sum_{i=0}^{L - 1} & \left[S \frac{N}{P} + 2^{i} \left( \alpha + \frac{N}{P} \beta \right) \right]                                             \\
                     & = S \frac{N}{P} \log_2 \left(\frac{P}{M} \right) + \left(\frac{P}{M} - 1\right) \left( \alpha + \frac{N}{P} \beta \right) \\
                     & \approx S \frac{N}{P} \log_2 \left(\frac{P}{M} \right) + \frac{P}{M} \alpha + \frac{N}{M} \beta.
\end{align*}

Second, consider the grouped merge phase.
This phase gathers the $k$NNGs from all ranks in the same group, including the local rank, concatenates them, and then optimizes the resulting graph for ANN search.
The communication cost of this operation using point-to-point operations is
\[
  \frac{P}{M} \cdot \left(\alpha + \frac{N}{P} \beta \right) = \frac{P}{M} \alpha + \frac{N}{M} \beta.
\]

Third, consider the flat refinement phase after grouped merge.
In this phase, each rank performs ANN searches for its $\frac{N}{P}$ local points on the $M-1$ search graphs associated with the other groups, each of which has its data split across $\frac{P}{M}$ ranks.
The total runtime of this phase is therefore
\[
  (M - 1) \left[ S \frac{N}{P} + \frac{P}{M} \left( \alpha + \frac{N}{P} \beta \right) \right] \approx S M \frac{N}{P} + P \alpha + N \beta.
\]

Finally, combining the complexities for the binary-tree-structured refinement, grouped merge, and flat refinement phases, we have an overall complexity of
\[
  S \frac{N}{P} \left( \log_2 \left(\frac{P}{M}\right) + M \right) + \alpha \left( P + \frac{2P}{M} \right) + \beta \left( N + \frac{2N}{M} \right).
\]

From this expression, we see that the computational requirements decrease as $P$ increases, as desired, and if the search cost scales with dataset size (for example, if $S \propto \log(\frac{N}{P})$), then the observed scalability may be better than predicted.
The communication costs show a latency term that increases with $P$ and a bandwidth term that remains constant for fixed values of $M$, both coming from a rank's need to retrieve data from all other ranks.
While these communication terms ultimately limit scalability, we expect the algorithm to scale well in practice over a broad range of $P$ because ANN search has high arithmetic intensity.
For example, a single query may visit $k^2$ neighbors within two hops of the initial entry point, so transferred data are reused extensively.
This reuse helps amortize communication cost, and we further mitigate its impact by overlapping communication and computation.
Consistent with this analysis, we do not observe these bottlenecks in the results presented in Section~\ref{sec:evaluation}.






\section{Evaluation}
\label{sec:evaluation}

\subsection{Hardware and Software Environment}

We use the Tuolumne cluster at Lawrence Livermore National Laboratory (LLNL) for our evaluation.
Tuolumne\footnote{https://hpc.llnl.gov/hardware/compute-platforms/tuolumne} is a sibling system of El Capitan, shares the same architecture, and was ranked 12th on the TOP500 list as of November 2025~\cite{TOP500}.
Each node of Tuolumne is equipped with four AMD MI300A APUs with 512 GB of memory in total and HPE Slingshot 11 interconnect.
We use ROCm v6.4.3.

We set the environment variable \texttt{HSA\_XNACK} to \texttt{1} to leverage the unified memory mode of the MI300A architecture, which allows us to avoid explicit data copies between CPU and GPU during our experiments.

\subsection{Datasets}

We consider 11 datasets, listed in Table~\ref{tbl:dataset}.
The top six datasets are from ANN-Benchmarks~\cite{ANNBenchmarks}, and the billion-scale base datasets are from the Big ANN Benchmarks~\cite{BigANNBenchmarks}.

Following prior studies, we use the first 100 million points of the DEEP-1B and SIFT-1B datasets as separate datasets, which we call \textit{DEEP-100M} and \textit{SIFT-100M}, respectively.
\textit{DEEP-2B} is a synthetic dataset generated from DEEP-1B to evaluate scalability beyond one billion data points.
We form two copies and shift them along different coordinate axes so their ranges do not overlap: in each copy, we add a scalar to one coordinate so the new minimum exceeds the original maximum by $\epsilon$.
We chose this simple construction because it keeps local data densities reasonably natural for ${L_2}$ distance.


\begin{table}
  \caption{Datasets used in our evaluation. \emph{Dims}:\ feature vector dimensions. \emph{Type}:\ feature vector element type.}
  \label{tbl:dataset}
  \centering
  \begin{tabular}{lrrcc}
    \toprule
    \multicolumn{1}{c}{Dataset}        & \multicolumn{1}{c}{Dims}       &
    \multicolumn{1}{c}{Points}         & \multicolumn{1}{c}{Similarity} &
    \multicolumn{1}{c}{Type}                                                                               \\
    \midrule
    Fashion-MNIST~\cite{Fashion-MNIST} & 784                            & 60,000      & ${L_2}$    & float \\
    NYTimes~\cite{NYTimes}             & 256                            & 290,000     & ${cosine}$ & float \\
    Last.fm~\cite{ANNBenchmarks}       & 65                             & 292,385     & ${cosine}$ & float \\
    GIST~\cite{SIFT1M_PQ_GIST}         & 960                            & 1,000,000   & ${L_2}$    & float \\
    SIFT-1M~\cite{SIFT1M_PQ_GIST}      & 128                            & 1,000,000   & ${L_2}$    & float \\
    GloVe50~\cite{GloVe}               & 50                             & 1,183,514   & ${cosine}$ & float \\
    \midrule
    DEEP-100M                          & 96                             & 100 million & ${L_2}$    & float \\
    SIFT-100M                          & 128                            & 100 million & ${L_2}$    & uint8 \\
    DEEP-1B~\cite{Deep1B}              & 96                             & 1 billion   & ${L_2}$    & float \\
    SIFT-1B~\cite{BigANN}              & 128                            & 1 billion   & ${L_2}$    & uint8 \\
    DEEP-2B                            & 96                             & 2 billion   & ${L_2}$    & float \\
    \bottomrule
  \end{tabular}
\end{table}

\subsection{Single APU $k$NNG Construction Performance}
\label{sec:single-apu-knng-construction}

We first evaluate the performance of our local lock-free $k$NNG construction algorithm on a single APU\@.

\subsubsection{Experimental Setup}
For performance comparison, we use the NN-Descent algorithm implemented in the hipVS library~\cite{hipVS},
which is based on a GPU-based NN-Descent algorithm, GNND~\cite{GPU-NNDescent-Wang}.

When evaluating neighbor graph construction, graph quality must be considered because it trades off against construction time: in general, higher recall requires more time.
We use recall@32 as the metric for graph quality.
Thus, we extract only the nearest 32 neighbors for each point from a constructed $k$NNG ($k \geq 32$) and count how many of them exist in the ground truth data computed by a brute-force method.
We measured only the graph construction time (e.g., excluding data loading time).
Specifically, we loaded the dataset into hipMalloc-allocated memory and passed a pointer to the graph construction functions.

For both implementations, we fixed the maximum number of NN-Descent iterations to 100 and varied the $k$ (number of neighbors).
To increase $k$NNG quality, it is common to use a $k$ value larger than the target number of neighbors and truncate the graph after construction (we used $k=32, 48, 64$).
We also varied other parameters such as $\delta$ (the convergence threshold) to explore the trade-off between recall and construction time.
We used several values between 0.00001 and 0.008 for $\delta$.
SOLANET also supports another parameter, $\rho$ (sampling rate), from NN-Descent, and we used 0.6, 0.8, and 1.0 (no sampling).

\subsubsection{Results}

\begin{figure}[t]
  \centering
  \begin{subfigure}[t]{0.49\columnwidth}
    \centering
    \includegraphics[width=\linewidth]{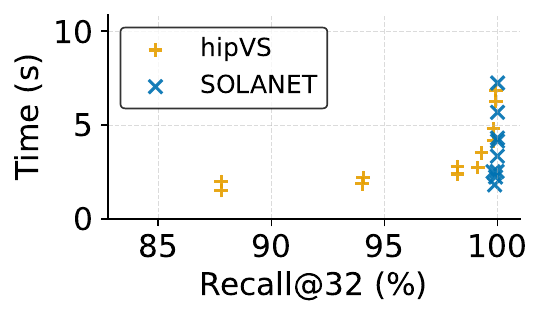}
    \caption{Fashion-MNIST}
  \end{subfigure}
  \hfill
  \begin{subfigure}[t]{0.49\columnwidth}
    \centering
    \includegraphics[width=\linewidth]{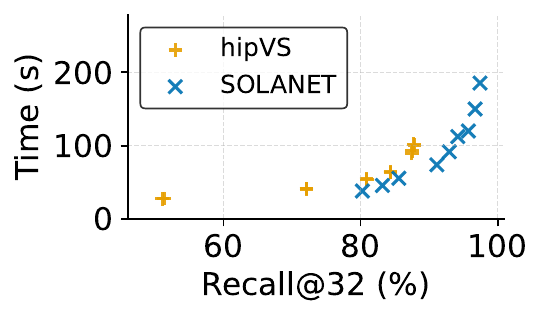}
    \caption{GIST}
  \end{subfigure}

  \vspace{0.5em}

  \begin{subfigure}[t]{0.49\columnwidth}
    \centering
    \includegraphics[width=\linewidth]{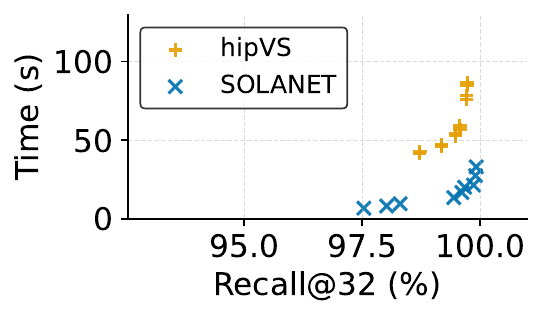}
    \caption{SIFT-1M}
  \end{subfigure}
  \hfill
  \begin{subfigure}[t]{0.49\columnwidth}
    \centering
    \includegraphics[width=\linewidth]{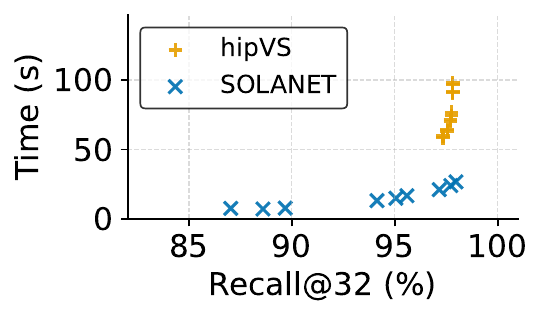}
    \caption{GloVe50}
  \end{subfigure}

  \vspace{0.5em}

  \begin{subfigure}[t]{0.49\columnwidth}
    \centering
    \includegraphics[width=\linewidth]{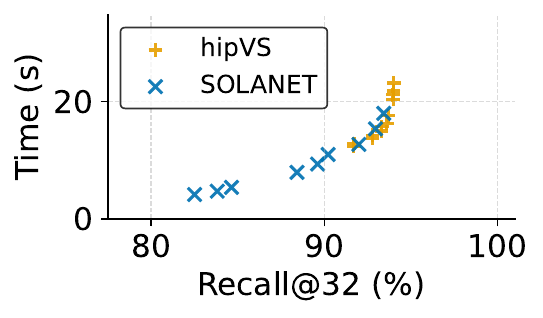}
    \caption{NYTimes}
  \end{subfigure}
  \hfill
  \begin{subfigure}[t]{0.49\columnwidth}
    \centering
    \includegraphics[width=\linewidth]{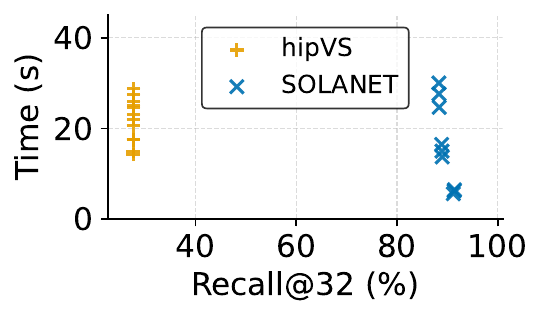}
    \caption{Last.fm}
  \end{subfigure}
  \caption{Single APU $k$NNG construction recall-time trade-off across six datasets. Each point corresponds to a specific combination of parameters.}
  \label{fig:single-build-speed-vs-recall}
\end{figure}

We show the recall@32 versus performance trade-off in Fig~\ref{fig:single-build-speed-vs-recall}, where each point corresponds to a specific parameter combination.
We present only the points within reasonable recall and construction time ranges for better visibility.

Overall, SOLANET's implementation achieved better recall-time trade-off than the hipVS implementation across all datasets except NYTimes, where hipVS performed slightly better.

On the Last.fm dataset, hipVS could not achieve more than a 28\% recall within the configuration space we explored, while SOLANET achieved around 91\% recall.
The ground-truth neighbor distances are small in the Last.fm dataset (e.g., less than 1.e-04).
We hypothesize that hipVS's low recall was due to floating-point precision issues or design choices that are not well-suited for handling such small distance values.

\subsection{Distributed $k$NNG Construction Performance}

We evaluate SOLANET on datasets with at least 100 million data points, focusing on strong scaling.

\subsubsection{Experimental Setup}

We constructed a $k$NNG with $k=32$ for the DEEP-100M and SIFT-100M datasets and $k=20$ for the DEEP-1B, SIFT-1B, and DEEP-2B datasets.
We report graph construction time excluding data loading time.
We set $\delta=0.0001$ and $\rho=0.5$ for all datasets.
We set $M$ to 2 for the 100-million-point datasets, 32 for the 1-billion-point datasets, and 64 for the 2-billion-point dataset.

We used 4 APUs per node and up to 128 nodes, depending on the dataset size.

For performance comparison, we also ran NEO-DNND~\cite{NEO-DNND}, which is a state-of-the-art distributed NN-Descent implementation for CPU clusters.
For NEO-DNND, we used a CPU-only cluster in which each node has two CPUs (64 cores per CPU) and 256 GB of DRAM\footnote{https://hpc.llnl.gov/hardware/compute-platforms/dane}.
We set $k=20$, $\delta=0.0001$, and $\rho=0.5$ for NEO-DNND\@.

\subsubsection{Scalability Analysis}

\begin{figure*}[t]
  \centering
  \begin{subfigure}[t]{0.31\textwidth}
    \centering
    \includegraphics[width=\linewidth]{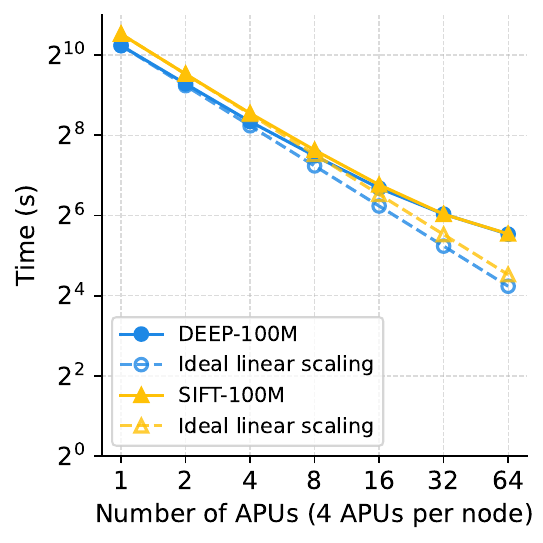}
    \caption{DEEP-100M and SIFT-100M datasets}
  \end{subfigure}
  \hfill
  \begin{subfigure}[t]{0.31\textwidth}
    \centering
    \includegraphics[width=\linewidth]{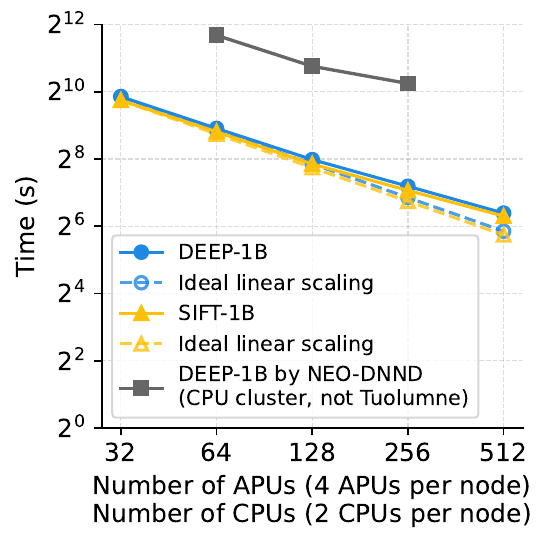}
    \caption{DEEP-1B and SIFT-1B datasets}
  \end{subfigure}
  \hfill
  \begin{subfigure}[t]{0.31\textwidth}
    \centering
    \includegraphics[width=\linewidth]{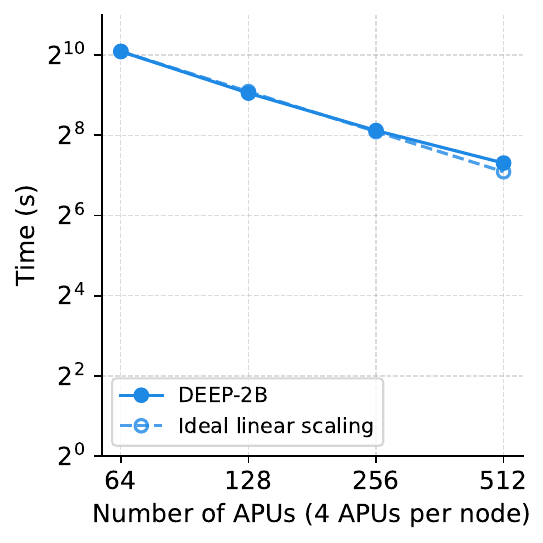}
    \caption{DEEP-2B dataset}
  \end{subfigure}
  \caption{Strong scaling study results by SOLANET on the Tuolumne cluster (4 AMD MI300A APUs per node). NEO-DNND~\cite{NEO-DNND} was run on the Dane (CPU-only) cluster for performance comparison. $k$NNG $k=20$.}
  \label{fig:strong-scaling}
\end{figure*}

We show the strong scaling results in Fig.~\ref{fig:strong-scaling}.

For the DEEP-100M and SIFT-100M datasets, increasing the number of APUs from 1 to 32 reduced the execution time from 1207 to 66 seconds (18.3x speedup) and from 1474 to 65 seconds (22.7x speedup), respectively.
By continuing to 64 APUs, it yeilded a 26x speedup for DEEP-100M and a 31.6x speedup for SIFT-100M\@.
Beyond 32 APUs, scalability began to degrade visibly.
We attribute this to the number of points per partition becoming too small to fully utilize the hardware and to increased communication overhead.

The DEEP-1B and SIFT-1B datasets required at least 32 APUs.
For the DEEP-1B dataset, increasing the number of APUs from 32 to 512 reduced the execution time from 926 seconds to 84 seconds, yielding an 11x speedup.
The SIFT-1B dataset showed a similar trend, with execution time decreasing from 858 seconds on 32 APUs to 79 seconds on 512 APUs, yielding an 11.7x speedup.

In comparison, NEO-DNND took 1,214 seconds on 256 CPUs to construct a $k$NNG with $k=20$ for the DEEP-1B dataset.
SOLANET completed the same task in 146 seconds on 256 APUs, achieving an 8.3x speedup over NEO-DNND\@.

For the DEEP-2B dataset, SOLANET required a minimum of 64 APUs to accommodate the dataset in memory.
Scaling from 64 to 512 APUs reduced the execution time from 1,090 seconds to 159 seconds, yielding a 6.9x speedup.

\subsubsection{Performance Breakdown}

\begin{figure}
  \centering
  \begin{subfigure}{\columnwidth}
    \centering
    \includegraphics[width=0.8\linewidth]{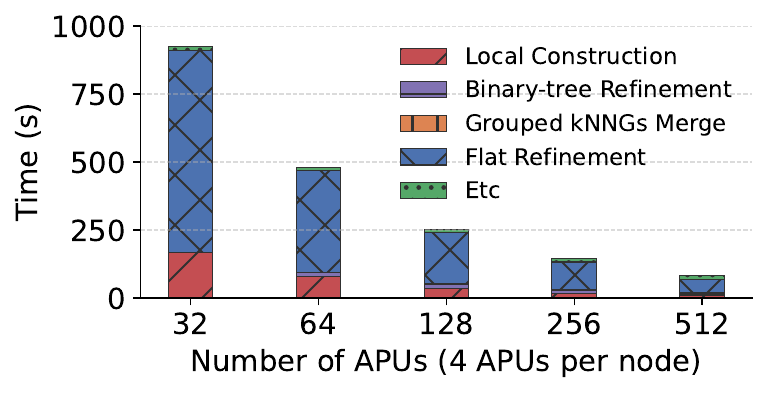}
    \caption{Absolute execution time.}
  \end{subfigure}

  \vspace{0.5em}

  \begin{subfigure}{\columnwidth}
    \centering
    \includegraphics[width=0.8\linewidth]{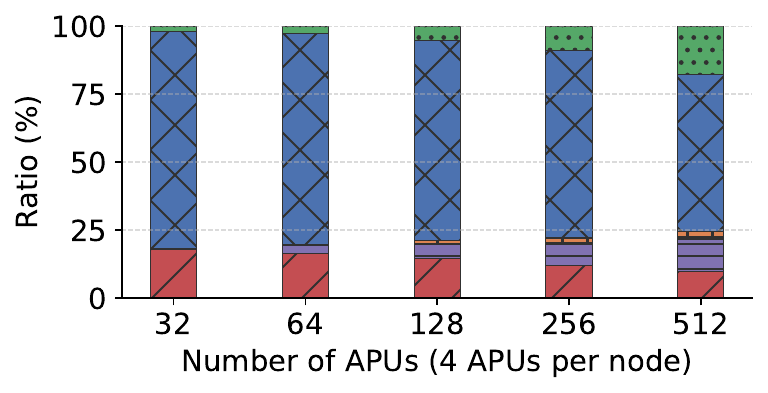}
    \caption{Performance breakdown ratio.}
  \end{subfigure}

  \caption{Performance breakdown of the distributed neighbor graph construction by SOLANET (DEEP-1B dataset).}
  \label{fig:strong-scaling-breakdown-deep1b}
\end{figure}

We show the performance breakdown for the DEEP-1B dataset in Fig.~\ref{fig:strong-scaling-breakdown-deep1b}.
Because other billion-scale datasets show very similar trends, we present only DEEP-1B.
We categorize the execution time into four categories: local $k$NNG construction, binary-tree-structured $k$NNG refinement, grouped $k$NNGs merge, and flat $k$NNG refinement.
Each category corresponds to a step in the distributed algorithm from Section~\ref{sec:dist-construction} and includes both computation and communication time.
The \textit{Etc} category includes other steps such as converting internal point IDs to external ones and deallocating workspace memory before completion.

First, the local $k$NNG construction time decreased from 166 seconds to 8 seconds (21x) as the number of APUs increased from 32 to 512.
Since the runtime of NN-Descent is superlinear in the number of data points,
the local $k$NNG construction time is expected to decrease superlinearly as the number of APUs increases.
We confirmed that our local NN-Descent implementation can achieve this expected behavior.

Next is the binary-tree-structured $k$NNG refinement step, which is a hierarchical merging process of sub-$k$NNGs.
This step merged sub-$k$NNGs into 32 sub-$k$NNGs, so it was not executed in the 32-APU case.
This step accounted for a small portion of the overall execution time.
For example, in the 512-APU case, which had the largest number of merges, merging 512 sub-$k$NNGs into 32 sub-$k$NNGs took 10 seconds and accounted for only 12\% of the total execution time.

The flat refinement phase took 739.89 seconds on 32 APUs and 48.68 seconds on 512 APUs (15x).
In both the 32-APU and 512-APU cases, each MPI rank ran ANN search against 31 sub-$k$NNGs on other ranks.
The number of query points per rank is 16 times smaller in the 512-APU case because the dataset is partitioned across 16 times more ranks.
We achieved a 15x speedup, which is close to the ideal scalability of 16x.

We also observed that CAGRA's local ANN search was highly robust to the number of source points in the search graph.
For example, it achieved 1.48 million QPS (queries per second) when conducting ANN searches on a $k$NNG with 2 million source points for 2 million query points simultaneously, and it achieved 1.3--1.4 million QPS on $k$NNGs with 32 million source points from the same query points.

We also observed that communication time during distributed graph refinement was negligible, owing to the communication optimizations, including one-sided communication, avoidance of fine-grained or irregular communication, and overlapping computation and communication when possible.


\subsubsection{Graph Quality Validation}
\label{sec:1b-knng-comparison}

To verify that SOLANET improves performance without degrading graph quality, we evaluated the constructed $k$NNGs in two settings.

For the small datasets in Section~\ref{sec:single-apu-knng-construction}, we built $k$NNGs with 1--16 APUs using the same parameters as in the billion-scale experiments except $k=32$ and $k_s=32$, and measured recall@32 against the ground truth.
SOLANET achieved 99\% recall@32 on all $L_2$ datasets regardless of APU count, while NYTimes stayed around 80\%, varying by only a few percentage points.
On Last.fm and GIST, recall@32 increased from 91\% to 95\% and from 86\% to 90\% when scaling from a single APU to 16 APUs, respectively.

For the billion-scale datasets, we compare SOLANET graphs with those produced by NEO-DNND~\cite{NEO-DNND}.
NN-Descent is known to produce high-quality graphs and NEO-DNND is intended to preserve that quality at scale.
We compute recall@20 by comparing a \emph{test} graph with a \emph{reference} graph using only neighbor distances: for each point, the distance to the 20th neighbor in the reference graph defines a threshold, and we count neighbors in the test graph within that threshold.
We ignore neighbor IDs because the reference graph is itself approximate (strictly speaking, this deviates from the exact definition of recall).
Using NEO-DNND graphs as the reference, SOLANET achieved 99\% recall@20 on both DEEP-1B and SIFT-1B for all APU counts.
Reversing the comparison, NEO-DNND achieved 70--71\% on DEEP-1B and 74--75\% on SIFT-1B.
This result indicates that SOLANET was able to construct $k$NNGs with substantially reduced runtime while maintaining high graph quality.
\section{Conclusion}
\label{sec:conclusion}

We presented SOLANET, a distributed neighbor graph construction toolkit for GPU-accelerated systems.

The evaluation shows that our algorithmic design is both efficient and scalable.
On a single MI300A APU, our local NN-Descent implementation outperforms a state-of-the-art GPU-based baseline while maintaining high graph quality.
On the Tuolumne supercomputer, SOLANET demonstrates strong scaling up to 512 APUs for billion-scale datasets, achieving significant reductions in end-to-end runtime while preserving high graph quality.

We believe SOLANET provides a practical and efficient solution for constructing large-scale neighbor graphs on distributed GPU systems, an increasingly important need as datasets continue to grow in size and dimensionality.

As future work, we plan to investigate additional ANN algorithms and libraries for local $k$NNG construction and ANN search, as well as to port SOLANET to NVIDIA GPUs.
Additionally, to avoid graph refinement between certain partitions and improve overall efficiency, we plan to explore partitioning strategies that leverage domain knowledge in specific applications.

\section*{Acknowledgment}
 {\small
  AI-based writing assistance tools were used throughout this manuscript just for text refinement with GPT-5~\cite{GPT5}. All AI-assisted text was carefully reviewed and verified by the authors. AI tools were not used to generate scientific results, data, or conclusions.
 }

\bibliographystyle{IEEEtran}
\bibliography{bib/IEEEabrv,bib/ref}

@STRING{jan = "Jan."}

@STRING{may = "May"}

@STRING{jun = "June"}

@STRING{jul = "July"}

@STRING{oct = "Oct."}

@STRING{dec = "Dec."}

@inproceedings{nndescent,
  address    = {New York, NY, USA},
  author     = {Dong, Wei and Moses, Charikar and Li, Kai},
  booktitle  = {Proceedings of the 20th International Conference on World Wide Web},
  doi        = {10.1145/1963405.1963487},
  isbn       = {9781450306324},
  location   = {Hyderabad, India},
  numpages   = {10},
  pages      = {577--586},
  publisher  = {Association for Computing Machinery},
  series     = {WWW '11},
  title      = {{E}fficient {K}-{N}earest {N}eighbor {G}raph {C}onstruction for {G}eneric {S}imilarity {M}easures},
  url        = {https://doi.org/10.1145/1963405.1963487},
  year       = {2011}
}

@inbook{GPU-NNDescent-Wang,
  address    = {New York, NY, USA},
  author     = {Wang, Hui and Zhao, Wan-Lei and Zeng, Xiangxiang and Yang, Jianye},
  booktitle  = {Proceedings of the 30th ACM International Conference on Information \& Knowledge Management},
  isbn       = {9781450384469},
  numpages   = {10},
  pages      = {1929--1938},
  publisher  = {Association for Computing Machinery},
  title      = {Fast K-NN Graph Construction by GPU Based NN-Descent},
  url        = {https://doi.org/10.1145/3459637.3482344},
  year       = {2021}
}

@misc{pynndescent,
  author       = {PyNNDescent},
  howpublished = {\url{https://github.com/lmcinnes/pynndescent}},
  note         = {[Accessed 24-Feb-2026]},
  title        = {{G}it{H}ub - lmcinnes/pynndescent: {A} {P}ython nearest neighbor descent for approximate nearest neighbors --- github.com}
}

@article{NSG,
  author     = {Fu, Cong and Xiang, Chao and Wang, Changxu and Cai, Deng},
  doi        = {10.14778/3303753.3303754},
  issn       = {2150-8097},
  issue_date = {January 2019},
  journal    = {Proc. VLDB Endow.},
  month      = {jan},
  number     = {5},
  numpages   = {14},
  pages      = {461--474},
  publisher  = {VLDB Endowment},
  title      = {Fast Approximate Nearest Neighbor Search with the Navigating Spreading-out Graph},
  url        = {https://doi.org/10.14778/3303753.3303754},
  volume     = {12},
  year       = {2019}
}

@article{ANNBenchmarks,
  author     = {Martin Aum{\"u}ller and Erik Bernhardsson and Alexander Faithfull},
  doi        = {https://doi.org/10.1016/j.is.2019.02.006},
  issn       = {0306-4379},
  journal    = {Information Systems},
  pages      = {101374},
  title      = {ANN-Benchmarks: A benchmarking tool for approximate nearest neighbor algorithms},
  url        = {https://doi.org/10.1016/j.is.2019.02.006},
  volume     = {87},
  year       = {2020}
}

@misc{BigANNBenchmarks,
  author       = {Harsha Vardhan Simhadri and George Williams and Martin Aum{\"u}ller and Artem Babenko and Dmitry Baranchuk and Qi Chen and Matthijs Douze and Lucas Hosseini and Ravishankar Krishnaswamy and Gopal Srinivasa and Suhas Jayaram Subramanya and Jingdong Wang},
  howpublished = {\url{http://big-ann-benchmarks.com/neurips21.html}},
  note         = {[Accessed 30-Mar-2026]},
  title        = {{B}illion-{S}cale {A}pproximate {N}earest {N}eighbor {S}earch {C}hallenge: {NeurIPS}'21 competition track}
}

@article{DiskANN,
  author   = {Krishnaswamy, Ravishankar and Krishnaswamy, Ravishankar and Manohar, M. and Simhadri, Harsha},
  title    = {The DiskANN library: Graph-Based Indices for Fast, Fresh and Filtered Vector Search},
  year     = {2024},
  month    = {December},
  url      = {{https://www.microsoft.com/en-us/research/publication/the-diskann-library-graph-based-indices-for-fast-fresh-and-filtered-vector-search/}},
  pages    = {20-42},
  journal  = {IEEE Data Eng. Bull.},
  volume   = {48}
}

@article{WARASHINA,
  author     = {Tomohiro WARASHINA and Kazuo AOYAMA and Hiroshi SAWADA and Takashi HATTORI},
  doi        = {10.1587/transinf.2014EDP7108},
  journal    = {IEICE Transactions on Information and Systems},
  number     = {12},
  pages      = {3142-3154},
  title      = {Efficient K-Nearest Neighbor Graph Construction Using MapReduce for Large-Scale Data Sets},
  url        = {https://doi.org/10.1587/transinf.2014EDP7108},
  volume     = {E97.D},
  year       = {2014}
}

@article{MapReduce,
  author    = {Dean, Jeffrey and Ghemawat, Sanjay},
  journal   = {Communications of the ACM},
  number    = {1},
  pages     = {107--113},
  publisher = {ACM New York, NY, USA},
  title     = {MapReduce: simplified data processing on large clusters},
  volume    = {51},
  year      = {2008},
  url       = {https://dl.acm.org/doi/10.1145/1327452.1327492}
}

@article{Wang,
  author     = {Wang, Mengzhao and Xu, Xiaoliang and Yue, Qiang and Wang, Yuxiang},
  doi        = {10.14778/3476249.3476255},
  issn       = {2150-8097},
  issue_date = {July 2021},
  journal    = {Proc. VLDB Endow.},
  month      = {jul},
  number     = {11},
  numpages   = {15},
  pages      = {1964--1978},
  publisher  = {VLDB Endowment},
  title      = {A Comprehensive Survey and Experimental Comparison of Graph-Based Approximate Nearest Neighbor Search},
  url        = {https://doi.org/10.14778/3476249.3476255},
  volume     = {14},
  year       = {2021}
}

@inproceedings{DNND,
  address    = {New York, NY, USA},
  author     = {Iwabuchi, Keita and Steil, Trevor and Priest, Benjamin and Pearce, Roger and Sanders, Geoffrey},
  booktitle  = {Proceedings of the SC '23 Workshops of The International Conference on High Performance Computing, Network, Storage, and Analysis},
  doi        = {10.1145/3624062.3625132},
  isbn       = {9798400707858},
  numpages   = {9},
  pages      = {730--738},
  publisher  = {Association for Computing Machinery},
  series     = {SC-W '23},
  title      = {Towards A Massive-scale Distributed Neighborhood Graph Construction},
  url        = {https://doi.org/10.1145/3624062.3625132},
  year       = {2023}
}

@inproceedings{GloVe,
  title     = {{G}lo{V}e: Global Vectors for Word Representation},
  author    = {Pennington, Jeffrey  and
               Socher, Richard  and
               Manning, Christopher},
  editor    = {Moschitti, Alessandro  and
               Pang, Bo  and
               Daelemans, Walter},
  booktitle = {Proceedings of the 2014 Conference on Empirical Methods in Natural Language Processing ({EMNLP})},
  month     = oct,
  year      = {2014},
  address   = {Doha, Qatar},
  publisher = {Association for Computational Linguistics},
  url       = {http://doi.org/10.3115/v1/D14-1162},
  doi       = {10.3115/v1/D14-1162},
  pages     = {1532--1543}
}

@inproceedings{Deep1B,
  author     = {Yandex, Artem Babenko and Lempitsky, Victor},
  booktitle  = {2016 IEEE Conference on Computer Vision and Pattern Recognition (CVPR)},
  doi        = {10.1109/CVPR.2016.226},
  pages      = {2055-2063},
  title      = {Efficient Indexing of Billion-Scale Datasets of Deep Descriptors},
  url        = {https://doi.org/10.1109/CVPR.2016.226},
  year       = {2016}
}

@inproceedings{BigANN,
  author     = {J{\'e}gou, Herv{\'e} and Tavenard, Romain and Douze, Matthijs and Amsaleg, Laurent},
  booktitle  = {2011 IEEE International Conference on Acoustics, Speech and Signal Processing (ICASSP)},
  doi        = {10.1109/ICASSP.2011.5946540},
  pages      = {861-864},
  title      = {Searching in one billion vectors: Re-rank with source coding},
  url        = {https://doi.org/10.1109/ICASSP.2011.5946540},
  year       = {2011}
}

@inproceedings{RAG,
  address   = {Red Hook, NY, USA},
  articleno = {793},
  author    = {Lewis, Patrick and Perez, Ethan and Piktus, Aleksandra and Petroni, Fabio and Karpukhin, Vladimir and Goyal, Naman and K\"{u}ttler, Heinrich and Lewis, Mike and Yih, Wen-tau and Rockt\"{a}schel, Tim and Riedel, Sebastian and Kiela, Douwe},
  booktitle = {Proceedings of the 34th International Conference on Neural Information Processing Systems},
  isbn      = {9781713829546},
  location  = {Vancouver, BC, Canada},
  numpages  = {16},
  publisher = {Curran Associates Inc.},
  series    = {NIPS'20},
  title     = {Retrieval-augmented generation for knowledge-intensive NLP tasks},
  year      = {2020},
  url       = {https://dl.acm.org/doi/abs/10.5555/3495724.3496517}
}

@inproceedings{Kim,
  author    = {Kim, Sang-Hong and Park, Ha-Myung},
  title     = {Efficient Distributed Approximate k-Nearest Neighbor Graph Construction by Multiway Random Division Forest},
  year      = {2023},
  isbn      = {9798400701030},
  publisher = {Association for Computing Machinery},
  address   = {New York, NY, USA},
  url       = {https://doi.org/10.1145/3580305.3599327},
  doi       = {10.1145/3580305.3599327},
  booktitle = {Proceedings of the 29th ACM SIGKDD Conference on Knowledge Discovery and Data Mining},
  pages     = {1097–1106},
  numpages  = {10},
  location  = {},
  series    = {KDD '23}
}

@inproceedings{CAGRA,
  author    = { Ootomo, Hiroyuki and Naruse, Akira and Nolet, Corey and Wang, Ray and Feher, Tamas and Wang, Yong },
  booktitle = { 2024 IEEE 40th International Conference on Data Engineering (ICDE) },
  title     = {{ CAGRA: Highly Parallel Graph Construction and Approximate Nearest Neighbor Search for GPUs }},
  year      = {2024},
  volume    = {},
  issn      = {},
  pages     = {4236-4247},
  doi       = {10.1109/ICDE60146.2024.00323},
  url       = {https://doi.ieeecomputersociety.org/10.1109/ICDE60146.2024.00323},
  publisher = {IEEE Computer Society},
  address   = {Los Alamitos, CA, USA},
  month     = May
}

@article{ADENIYI201690,
  title    = {Automated web usage data mining and recommendation system using K-Nearest Neighbor (KNN) classification method},
  journal  = {Applied Computing and Informatics},
  volume   = {12},
  number   = {1},
  pages    = {90-108},
  year     = {2016},
  issn     = {2210-8327},
  doi      = {https://doi.org/10.1016/j.aci.2014.10.001},
  url      = {https://doi.org/10.1016/j.aci.2014.10.001},
  author   = {D.A. Adeniyi and Z. Wei and Y. Yongquan}
}

@article{LIAO2002,
  title    = {Use of K-Nearest Neighbor classifier for intrusion detection11An earlier version of this paper is to appear in the Proceedings of the 11th USENIX Security Symposium, San Francisco, CA, August 2002},
  journal  = {Computers \& Security},
  volume   = {21},
  number   = {5},
  pages    = {439-448},
  year     = {2002},
  issn     = {0167-4048},
  doi      = {https://doi.org/10.1016/S0167-4048(02)00514-X},
  url      = {https://doi.org/10.1016/S0167-4048(02)00514-X},
  author   = {Yihua Liao and V.Rao Vemuri}
}

@article{MovieLens,
  author     = {Harper, F. Maxwell and Konstan, Joseph A.},
  title      = {The MovieLens Datasets: History and Context},
  year       = {2015},
  issue_date = {January 2016},
  publisher  = {Association for Computing Machinery},
  address    = {New York, NY, USA},
  volume     = {5},
  number     = {4},
  issn       = {2160-6455},
  url        = {https://doi.org/10.1145/2827872},
  doi        = {10.1145/2827872},
  journal    = {ACM Trans. Interact. Intell. Syst.},
  month      = {dec},
  articleno  = {19},
  numpages   = {19}
}

@article{GGNN,
  author   = {Groh, Fabian and Ruppert, Lukas and Wieschollek, Patrick and Lensch, Hendrik P. A.},
  journal  = {IEEE Transactions on Big Data},
  title    = {GGNN: Graph-Based GPU Nearest Neighbor Search},
  year     = {2023},
  volume   = {9},
  number   = {1},
  pages    = {267-279},
  doi      = {10.1109/TBDATA.2022.3161156},
  url      = {https://doi.org/10.1109/TBDATA.2022.3161156}
}

@inproceedings{Wahlgren2025,
  author    = {Wahlgren, Jacob and Schieffer, Gabin and Shi, Ruimin and León, Edgar A. and Pearce, Roger and Gokhale, Maya and Peng, Ivy},
  booktitle = {2025 IEEE International Symposium on Workload Characterization (IISWC)},
  title     = {Dissecting CPU-GPU Unified Physical Memory on AMD MI300A APUs},
  year      = {2025},
  volume    = {},
  number    = {},
  pages     = {368-380},
  doi       = {10.1109/IISWC66894.2025.00038},
  url       = {https://doi.org/10.1109/IISWC66894.2025.00038}
}

@misc{cuVS,
  howpublished = {\url{https://github.com/rapidsai/cuvs}},
  note         = {[Accessed 25-Mar-2026]},
  title        = {{cuVS: Vector Search and Clustering on the GPU}}
}

@misc{hipVS,
  howpublished = {\url{https://github.com/ROCm-DS/hipVS}},
  note         = {[Accessed 25-Mar-2026]},
  title        = {{hipVS: GPU-accelerated vector search for AMD GPUs}}
}

@online{Fashion-MNIST,
  author      = {Han Xiao and Kashif Rasul and Roland Vollgraf},
  title       = {Fashion-MNIST: a Novel Image Dataset for Benchmarking Machine Learning Algorithms},
  date        = {2017-08-28},
  year        = {2017},
  eprintclass = {cs.LG},
  eprinttype  = {arXiv},
  eprint      = {cs.LG/1708.07747},
  url         = {https://doi.org/10.48550/arXiv.1708.07747}
}

@article{SIFT1M_PQ_GIST,
  title       = {{Product Quantization for Nearest Neighbor Search}},
  author      = {J{\'e}gou, Herv{\'e} and Douze, Matthijs and Schmid, Cordelia},
  url         = {https://doi.org/10.1109/TPAMI.2010.57},
  journal     = {{IEEE Transactions on Pattern Analysis and Machine Intelligence}},
  publisher   = {{Institute of Electrical and Electronics Engineers}},
  volume      = {33},
  number      = {1},
  pages       = {117-128},
  year        = {2011},
  month       = Jan,
  doi         = {10.1109/TPAMI.2010.57},
  hal_id      = {inria-00514462},
  hal_version = {v2}
}

@misc{NYTimes,
  author       = {Newman, David},
  title        = {{Bag of Words}},
  year         = {2008},
  howpublished = {UCI Machine Learning Repository},
  note         = {{DOI}: https://doi.org/10.24432/C5ZG6P}
}

@inproceedings{NEO-DNND,
  author    = {Iwabuchi, Keita and Steil, Trevor and Priest, Benjamin W. and Pearce, Roger and Sanders, Geoffrey},
  booktitle = {SC24-W: Workshops of the International Conference for High Performance Computing, Networking, Storage and Analysis},
  title     = {NEO-DNND: Communication-Optimized Distributed Nearest Neighbor Graph Construction},
  year      = {2024},
  volume    = {},
  number    = {},
  pages     = {688-696},
  doi       = {10.1109/SCW63240.2024.00096},
  url       = {https://doi.org/10.1109/SCW63240.2024.00096}
}

@article{1704843,
  author   = {Franti, P. and Virmajoki, O. and Hautamaki, V.},
  journal  = {IEEE Transactions on Pattern Analysis and Machine Intelligence},
  title    = {Fast Agglomerative Clustering Using a k-Nearest Neighbor Graph},
  year     = {2006},
  volume   = {28},
  number   = {11},
  pages    = {1875-1881},
  doi      = {10.1109/TPAMI.2006.227},
  url      = {https://doi.org/10.1109/TPAMI.2006.227}
}

@article{BRITO199733,
  title    = {Connectivity of the mutual k-nearest-neighbor graph in clustering and outlier detection},
  journal  = {Statistics \& Probability Letters},
  volume   = {35},
  number   = {1},
  pages    = {33-42},
  year     = {1997},
  issn     = {0167-7152},
  doi      = {https://doi.org/10.1016/S0167-7152(96)00213-1},
  url      = {https://doi.org/10.1016/S0167-7152(96)00213-1},
  author   = {M.R. Brito and E.L. Chávez and A.J. Quiroz and J.E. Yukich}
}

@article{QIN20181,
  title    = {A Novel clustering method based on hybrid K-nearest-neighbor graph},
  journal  = {Pattern Recognition},
  volume   = {74},
  pages    = {1-14},
  year     = {2018},
  issn     = {0031-3203},
  doi      = {https://doi.org/10.1016/j.patcog.2017.09.008},
  url      = {https://doi.org/10.1016/j.patcog.2017.09.008},
  author   = {Yikun Qin and Zhu Liang Yu and Chang-Dong Wang and Zhenghui Gu and Yuanqing Li}
}

@inproceedings{Pan2024,
  author    = {Pan, James Jie and Wang, Jianguo and Li, Guoliang},
  title     = {Vector Database Management Techniques and Systems},
  year      = {2024},
  isbn      = {9798400704222},
  publisher = {Association for Computing Machinery},
  address   = {New York, NY, USA},
  url       = {https://doi.org/10.1145/3626246.3654691},
  doi       = {10.1145/3626246.3654691},
  booktitle = {Companion of the 2024 International Conference on Management of Data},
  pages     = {597–604},
  numpages  = {8},
  location  = {Santiago AA, Chile},
  series    = {SIGMOD '24}
}

@misc{UMAP,
  title         = {UMAP: Uniform Manifold Approximation and Projection for Dimension Reduction},
  author        = {Leland McInnes and John Healy and James Melville},
  year          = {2020},
  eprint        = {1802.03426},
  archiveprefix = {arXiv},
  primaryclass  = {stat.ML},
  url           = {https://doi.org/10.48550/arXiv.1802.03426}
}

@inproceedings{SONG,
  author    = {Zhao, Weijie and Tan, Shulong and Li, Ping},
  booktitle = {2020 IEEE 36th International Conference on Data Engineering (ICDE)},
  title     = {SONG: Approximate Nearest Neighbor Search on GPU},
  year      = {2020},
  volume    = {},
  number    = {},
  pages     = {1033-1044},
  doi       = {10.1109/ICDE48307.2020.00094},
  url       = {https://doi.org/10.1109/ICDE48307.2020.00094}
}

@article{BANG,
  author   = {Venkatasubba, Karthik and Khan, Saim and Singh, Somesh and Simhadri, Harsha Vardhan and Vedurada, Jyothi},
  journal  = {IEEE Transactions on Big Data},
  title    = {BANG: Billion-Scale Approximate Nearest Neighbour Search Using a Single GPU},
  year     = {2025},
  volume   = {11},
  number   = {6},
  pages    = {3142-3157},
  doi      = {10.1109/TBDATA.2025.3581085},
  url      = {https://doi.org/10.1109/TBDATA.2025.3581085}
}

@misc{liu2026,
  title         = {GPU-Accelerated Algorithms for Graph Vector Search: Taxonomy, Empirical Study, and Research Directions},
  author        = {Yaowen Liu and Xuejia Chen and Anxin Tian and Haoyang Li and Qinbin Li and Xin Zhang and Alexander Zhou and Chen Jason Zhang and Qing Li and Lei Chen},
  year          = {2026},
  eprint        = {2602.16719},
  archiveprefix = {arXiv},
  primaryclass  = {cs.DB},
  url           = {https://doi.org/10.48550/arXiv.2602.16719}
}

@article{Faiss,
  author   = {Douze, Matthijs and Guzhva, Alexandr and Deng, Chengqi and Johnson, Jeff and Szilvasy, Gergely and Mazaré, Pierre-Emmanuel and Lomeli, Maria and Hosseini, Lucas and Jégou, Hervé},
  journal  = {IEEE Transactions on Big Data},
  title    = {The Faiss Library},
  year     = {2026},
  volume   = {12},
  number   = {2},
  pages    = {346-361},
  doi      = {10.1109/TBDATA.2025.3618474},
  url      = {https://doi.org/10.1109/TBDATA.2025.3618474}
}

@inproceedings{RengaBashyam2020,
  author    = {Renga Bashyam, K G and Vadhiyar, Sathish},
  booktitle = {2020 IEEE International Conference on Cluster Computing (CLUSTER)},
  title     = {Fast Scalable Approximate Nearest Neighbor Search for High-dimensional Data},
  year      = {2020},
  volume    = {},
  number    = {},
  pages     = {294-302},
  doi       = {10.1109/CLUSTER49012.2020.00040},
  url       = {https://doi.org/10.1109/CLUSTER49012.2020.00040}
}

@article{PyRKNN,
  author     = {Ruys, William and Ghafouri, Ali and Chen, Chao and Biros, George},
  title      = {Scalable KNN Graph Construction for Heterogeneous Architectures},
  year       = {2025},
  issue_date = {September 2025},
  publisher  = {Association for Computing Machinery},
  address    = {New York, NY, USA},
  volume     = {12},
  number     = {3},
  issn       = {2329-4949},
  url        = {https://doi.org/10.1145/3733610},
  doi        = {10.1145/3733610},
  journal    = {ACM Trans. Parallel Comput.},
  month      = jun,
  articleno  = {8},
  numpages   = {35}
}

@article{hie2019efficient,
  title     = {Efficient integration of heterogeneous single-cell transcriptomes using Scanorama},
  author    = {Hie, Brian and Bryson, Bryan and Berger, Bonnie},
  journal   = {Nature biotechnology},
  volume    = {37},
  number    = {6},
  pages     = {685--691},
  year      = {2019},
  publisher = {Nature Publishing Group US New York},
  url       = {https://doi.org/10.1038/s41587-019-0113-3}
}

@article{wolf2018scanpy,
  title     = {SCANPY: large-scale single-cell gene expression data analysis},
  author    = {Wolf, F Alexander and Angerer, Philipp and Theis, Fabian J},
  journal   = {Genome biology},
  volume    = {19},
  number    = {1},
  pages     = {15},
  year      = {2018},
  publisher = {Springer},
  url       = {https://doi.org/10.1186/s13059-017-1382-0}
}

@article{dries2021giotto,
  title     = {Giotto: a toolbox for integrative analysis and visualization of spatial expression data},
  author    = {Dries, Ruben and Zhu, Qian and Dong, Rui and Eng, Chee-Huat Linus and Li, Huipeng and Liu, Kan and Fu, Yuntian and Zhao, Tianxiao and Sarkar, Arpan and Bao, Feng and others},
  journal   = {Genome biology},
  volume    = {22},
  number    = {1},
  pages     = {78},
  year      = {2021},
  publisher = {Springer},
  url       = {https://doi.org/10.1186/s13059-021-02286-2}
}

@article{van2014accelerating,
  author     = {Van Der Maaten, Laurens},
  title      = {Accelerating t-SNE using tree-based algorithms},
  year       = {2014},
  issue_date = {January 2014},
  publisher  = {JMLR.org},
  volume     = {15},
  number     = {1},
  issn       = {1532-4435},
  journal    = {J. Mach. Learn. Res.},
  month      = jan,
  pages      = {3221–3245},
  numpages   = {25},
  url        = {https://dl.acm.org/doi/10.5555/2627435.2697068}
}

@article{Hockney1994,
  title    = {The communication challenge for {MPP}: {I}ntel {P}aragon and {M}eiko {CS-2}},
  journal  = {Parallel Computing},
  volume   = {20},
  number   = {3},
  pages    = {389-398},
  year     = {1994},
  issn     = {0167-8191},
  doi      = {https://doi.org/10.1016/S0167-8191(06)80021-9},
  url      = {https://doi.org/10.1016/S0167-8191(06)80021-9},
  author   = {Roger W. Hockney}
}

@misc{TOP500,
  author       = {TOP500.org},
  howpublished = {\url{https://top500.org/}},
  note         = {[Accessed 20-Mar-2026]},
  title        = {{TOP500 list}}
}

@misc{GPT5,
  author       = {OpenAI},
  howpublished = {\url{https://openai.com/gpt-5/}},
  note         = {[Accessed 5-Apr-2026]},
  title        = {{GPT5}}
}

\end{document}